\documentclass[a4paper,11pt]{article}
\usepackage{aaskaiid}
\usepackage{threeparttable}
\usepackage[encapsulated]{CJK}

\usepackage{subcaption}
\usepackage{orcidlink}

\title{Gravitational Light Deflection with SKA-VLBI and Its Application to Precision Tests of General Relativity}
\ShortTitle{Light Deflection}

\author[1]{Li, Y. J.\orcidlink{0000-0001-7526-0120}}
\ShortName{Li et al.} 
\author[1]{Li, J. J.\orcidlink{0000-0002-3338-8465}}
\author[1]{Lin, Z. H.\orcidlink{0009-0006-0392-6345}}
\author[2,3]{Liu, D. J.\orcidlink{0009-0001-9837-9455}}
\author[1]{Dong, Y. W.\orcidlink{0009-0000-5512-9109}}
\author[1]{Hao, C. J.\orcidlink{0000-0002-6820-198X}}
\author[1]{$^{\ast}$Xu, Y.\orcidlink{0000-0001-5602-3306}}

\affiliation[1]{Purple Mountain Observatory, Chinese Academy of Sciences, Nanjing 210023, China}
\emailAdd{xuye@pmo.ac.cn, liyj@pmo.ac.cn, linzh@pmo.ac.cn}
\affiliation[2]{College of Science, China Three Gorges University, Yichang 443002, China}
\affiliation[3]{Center for Astronomy and Space Sciences, China Three Gorges University, Yichang 443002, China}

\abstract{Experimental test of general relativity remains an ongoing endeavour. Radio astrometry provides a vital tool for precisely measuring the light deflection caused by the Sun, testing general relativity, and discriminating between gravitational theories. 
The best accuracy for the post-Newtonian relativistic parameter, $\gamma$, achieved with very long baseline interferometry is $9 \times 10^{-5}$. With 300-sec integration, SKA-VLBI can achieve a sensitivity of $\sim$15 $\mu$Jy at 15 GHz over a bandwidth of 0.256 GHz. This enables detection of $\sim$36 extragalactic radio sources per square degree with flux densities of $\sim$1.5 mJy, and potentially detecting in-beam radio sources. Single-epoch SKA-VLBI observations may achieve an astrometric precision of $\sim$2 $\mu$as. Utilising the Sun as a gravitational lens, 10-epoch positional tracking of extragalactic sources could improve $\gamma$ accuracy to $\sim$10$^{-7}$. Even with Jupiter as a lens, SKA-VLBI can measure $\gamma$ to $\sim$10$^{-4}$. Critically, it may conduct the first measurement of quadrupolar deflection of light caused by Jupiter, determining the physical oblateness of Jupiter, $J_{\mathrm{2, J}}$, to within $\sim${}$10^{-3}$. These advances are expected to rigorously test and improve gravitational theories or high-order parameterized post-Newtonian formalisms, while laying the foundations for (sub)$\mu$as astrometry.}

\begin{document}
\newcommand{\actaa}{Acta Astron.} 
\newcommand{\araa}{Annu. Rev. Astron. Astrophys.} 
\newcommand{\aar}{Astron. Astrophys. Rev.} 
\newcommand{\ab}{Astrobiol.} 
\newcommand{\aj}{Astron. J.} 
\newcommand{\apj}{Astrophys. J.} 
\newcommand{\apjl}{Astrophys. J. Lett.} 
\newcommand{\apjs}{Astrophys. J. Suppl. Ser.} 
\newcommand{\ao}{Appl. Opt.} 
\newcommand{\apss}{Astrophys. Space Sci.} 
\newcommand{\aap}{Astron. Astrophys.} 
\newcommand{\aapr}{Astron. Astrophys. Rev.} 
\newcommand{\aaps}{Astron. Astrophys. Suppl.} 
\newcommand{\baas}{Bull. Am. Astron. Soc.} 
\newcommand{\caa}{Chinese Astron. Astrophys.} 
\newcommand{\cjaa}{Chinese J. Astron. Astrophys.} 
\newcommand{\cqg}{Class. Quantum Gravity} 
\newcommand{\gal}{Galaxies} 
\newcommand{\gca}{Geochim. Cosmochim. Acta} 
\newcommand{\icarus}{Icarus} 
\newcommand{\jcap}{J. Cosmol. Astropart. Phys.} 
\newcommand{\jgr}{J. Geophys. Res.} 
\newcommand{\jgrp}{J. Geophys. Res.: Planets} 
\newcommand{\jqsrt}{J. Quant. Spectrosc. Radiat. Transf.} 
\newcommand{\memsai}{Mem. Soc. Astron. Italiana} 
\newcommand{\mnras}{Mon. Not. R. Astron. Soc.} 
\newcommand{\nat}{Nature} 
\newcommand{\nastro}{Nat. Astron.} 
\newcommand{\ncomms}{Nat. Commun.} 
\newcommand{\nphys}{Nat. Phys.} 
\newcommand{\na}{New Astron.} 
\newcommand{\nar}{New Astron. Rev.} 
\newcommand{\physrep}{Phys. Rep.} 
\newcommand{\pra}{Phys. Rev. A} 
\newcommand{\prb}{Phys. Rev. B} 
\newcommand{\prc}{Phys. Rev. C} 
\newcommand{\prd}{Phys. Rev. D} 
\newcommand{\pre}{Phys. Rev. E} 
\newcommand{\prl}{Phys. Rev. Lett.} 
\newcommand{\psj}{Planet. Sci. J.} 
\newcommand{\planss}{Planet. Space Sci.} 
\newcommand{\pnas}{Proc. Natl Acad. Sci. USA} 
\newcommand{\procspie}{Proc. SPIE} 
\newcommand{\pasa}{Publ. Astron. Soc. Aust.} 
\newcommand{\pasj}{Publ. Astron. Soc. Jpn} 
\newcommand{\pasp}{Publ. Astron. Soc. Pac.} 
\newcommand{\rmxaa}{Rev. Mexicana Astron. Astrofis.} 
\newcommand{\sci}{Science} 
\newcommand{\sciadv}{Sci. Adv.} 
\newcommand{\solphys}{Sol. Phys.} 
\newcommand{\sovast}{Soviet Ast.} 
\newcommand{\ssr}{Space Sci. Rev.} 
\newcommand{\uni}{Universe} 
\newcommand{\prx}{Phys. Rev. X} 
\newcommand{\lrr}{Living Rev. Relativ.} %
\newcommand{\raa}{Res. Astron. Astrophys.} %
\newcommand{\pla}{Phys. Lett. A} %
\maketitle

\begin{CJK*}{UTF8}{gbsn}
	
\section{Introduction}\label{sec:introduction-light-deflection}

Since the birth of general relativity (GR) in 1915, experimental test has remained an ongoing endeavour. The 1919 measurement of light deflection by the Sun, which verified GR at the sub-arcsecond level, was heralded as a ``Revolution in science/new theory of the universe/Newtonian ideas overthrown."
Light deflection and the theories behind it, including relativistic gravity and gravitational lensing theory, have emerged as pivotal tools in astronomy and cosmology \citep{Will2015}. Modern measurements of light deflection, alongside other weak-field tests, are commonly characterized by the parameterized post-Newtonian (PPN) formalism. This framework employs dimensionless parameters such as $\gamma$, $\beta$, $\xi$ to parameterize weak-field and slow-motion approximations in gravitational theories.  In GR, $\gamma = \beta = 1$ and all other parameters are null \citep{Will2014}. It is $\gamma$ that characterizes the impact of the scalar field on light propagation, quantifying the space curvature per unit mass.

\citet{Will2014} summarised light deflection measurements and the corresponding $\gamma$ accuracies. The most accurate measurement of $\gamma$ achieves an accuracy of $\sim${}$2\times 10^{-5}$, obtained through Doppler tracking of the Cassini spacecraft on the way to Saturn \citep{Bertotti+2003}. High-precision radio astrometry has become essential for precisely measuring light deflection by solar system objects and testing GR. Utilising the Sun as a gravitational lens, Very Long Baseline Interferometry (VLBI) measurements of $\gamma$ have progressed in accuracy from $\sim${}$5\times 10^{-3}$ \citep{Robertson-Carter1984} to $\sim${}$9\times 10^{-5}$ \citep{Titov+2018}. Only a few studies used Jupiter as the lens to measure $\gamma$: through Very Long Baseline Array observations, \citet{Fomalont-Kopeikin2008} obtained $\gamma = 1.01 \pm 0.03$, while \citet{Li+2022} yielded $\gamma = 0.984 \pm 0.037$.

Light deflection reaches $\sim${}$5\times10^5$ $\mu$as for paths passing at 1$^{\circ}$ from the Sun, while the maximum deflection induced by Jupiter is $\sim${}$1.6\times10^4$ $\mu$as. Light deflections caused by other solar system objects may also affect high-precision astrometry \citep{Li+2022b}. 
The light deflection effect, whose importance to astrometry was recognized in the 1980s and which was integrated into VLBI data processing pipeline, has been fundamentally linked with high-precision astrometry. 

Current parallax accuracies of masers in high-mass star-forming regions reach $\sim$6 $\mu$as \citep{Bian+2024}. Determination of possible spiral arm bifurcation points in the Milky Way's first quadrant requires $\lesssim$2 $\mu$as parallax accuracy \citep[see][in this book]{XuYe01.2026.SKA}. Meanwhile, quadrupolar deflection of light caused by Jupiter, which is not incorporated into the GR model of VLBI data process pipeline, can reach an order of ten $\mu$as \citep{Kopeikin-Makarov2007}. 
These necessitate advancing comprehension of (high-order) light deflection effects and optimising their calibration accuracy.

The SKA-VLBI will enhance the sensitivity of current VLBI by approximately one order of magnitude (reaching $\mu$Jy levels) and advance astrometric precision at $\mu$as-scales, potentially achieving 1 $\mu$as \citep{Paragi+2015, Li+2024-SKA-VLBI}. These enhancements will significantly improve $\gamma$ accuracy via light deflection measurements, and enable detection of multipolar deflection of light by solar system planets, thus facilitating rigorous gravitational physics tests and theoretical advances. This work presents an analysis of light deflection measurements with SKA-VLBI, the achievable accuracy of $\gamma$, its scientific applications, and potential breakthroughs.

\section{Specifications of SKA-VLBI}\label{sec:specification-light-deflection}

We focus on the continuum observations at 15 and 8 GHz with SKA-VLBI. Table \ref{tab:SKA-specification-light-deflection} presents the sensitivity, $\Delta S$, for SKA-VLBI with a 0.256 GHz bandwidth, 300-sec integration and 10,000 km baseline, based on the calculation in \citet{Li+2024-SKA-VLBI}, in which possible SKA-VLBI configurations are also presented. The position measurement uncertainty (i.e., astrometric
precision) introduced by thermal noise reads
\begin{equation}
    \Delta \theta_{\mathrm{thermal}} \approx 0.5 \frac{\theta_{\mathrm{beam}}}{\mathrm{DR}}.
\end{equation}
where $\theta_{\mathrm{beam}}$ and DR are the full width at half maximum of the beam and the dynamic range of the map, respectively (see Table \ref{tab:SKA-specification-light-deflection}). 

\begin{table*}[htbp]
\vspace{-0.2cm}
	\centering
		\footnotesize
		\setlength\tabcolsep{5pt}
	\renewcommand{\arraystretch}{1.1}
	\begin{threeparttable}
		\caption{Characterizing SKA-VLBI Astrometry \label{tab:SKA-specification-light-deflection}}
		\begin{tabular}{ccccccccc}
			\hline \hline
			Frequency & Baseline & Integration time & Bandwidth & $\Delta S$ & $\theta_{\mathrm{beam}}$ & DR & $\Delta \theta$ & $n_{\mathrm{ERS}}$ \\
			(GHz) &  (km) & (sec) & (GHz) & ($\mu$Jy beam$^{-1}$) & ($\mu$as) & & ($\mu$as) & (deg$^{-2}$)\\
			\hline
			  8.0  & $\sim$10,000 & 300 & 0.256 & $\sim$10 & $\sim$800 & 100:1 & $\sim$4 & $\sim$39\\
			15.0 & $\sim$10,000 & 300 & 0.256 & $\sim$15 & $\sim$400 & 100:1 & $\sim$2 & $\sim$36\\     
			\hline
		\end{tabular}
	\end{threeparttable}
\end{table*}

The solar corona or Jovian atmosphere would introduce systematic errors \citep[see][]{Kopeikin-Makarov2007, Li+2022}. For instance, at 15 GHz, the deflection caused by Jovian atmosphere is $\alpha_{\mathrm{pla, J}} \sim 16$--1.0 $\mu$as for rays passing at angular separations of $\chi = 1'$--$2'$, assuming a power-law index of $A = 2$ \citep[see][]{Kopeikin-Makarov2007}. For the Sun, $\alpha_{\mathrm{pla, S}} \sim 5,000$--500 $\mu$as at $\chi = 1^{\circ}$--$3^{\circ}$ \citep{Li+2022}. These atmospheric effects on astrometric precision, which are dominated by plasma, are analogous to those of Earth's ionosphere. System errors from both the ionospheric and tropospheric effects scale proportionally with the angular separation of source pairs \citep{Rioja-Dodson2020}. The expected density of extragalactic radio sources, $n_{\mathrm{ERS}}$, at 1.5 mJy is $\sim$36 deg$^{-2}$ (see Table \ref{tab:SKA-specification-light-deflection}) based on the simulation in \citet{Bonaldi+2021}. Therefore, in-beam observation could potentially be conducted because $\sim$50\% of them will be separated by less than the 15-m SKA-Mid primary beam assuming random distribution. The improved MultiView method can further reduce the systematic errors introduced by ionospheric and tropospheric residuals to $\sim$1 $\mu$as \citep{Rioja-Dodson2020}. Thanks to the ultra-high sensitivity of SKA Mid, phase information can be found at 1.5 mJy with 1-sec integration, enabling short-timescale phase variations to be traced. Consequently, effects from the atmosphere of the Sun or Jupiter may be addressed using shorter cycling intervals to mitigate phase jump errors. Multi-frequency scheduling (e.g., at 15 and 8 GHz) or wide-bandwidth observations (e.g., $\sim$5 GHz), together with plasma model fits, may further reduce the aforementioned plasma effects \citep{Kopeikin-Makarov2007, Fomalont+2009}.

The radio emission environment near Jupiter (e.g., at $1'$) or the Sun (e.g., at $1^{\circ}$) would introduce systematic noise into the maps. The phase of those sources may be resolved using the phase of sources at larger angular separations. This approach can reduce systematic errors introduced by the radio emission from Jupiter or the Sun. Systematic errors due to source structure can be effectively mitigated by selecting compact sources.

With the strategies outlined above, systematic errors may be constrained to a level comparable to the thermal noise.  The total astrometric precision, $\Delta \theta$, is therefore ideally assumed to be $\Delta \theta_{\mathrm{thermal}}$ with DR of 100:1 (see Table \ref{tab:SKA-specification-light-deflection}). Note that the actual precision will ultimately depend on real observations. 

\section{Gravitational Deflection of Light in the Solar System}\label{sec:light-deflection-solar-light-deflection}

\subsection{Expected Accuracy of \texorpdfstring{$\gamma$}{gamma} via Monopolar Light Deflection}\label{sec:light-deflection-monopolar-light-deflection}

\begin{figure}[ht]
    \centering
    \subfloat[]{\includegraphics[height=0.2\textwidth]{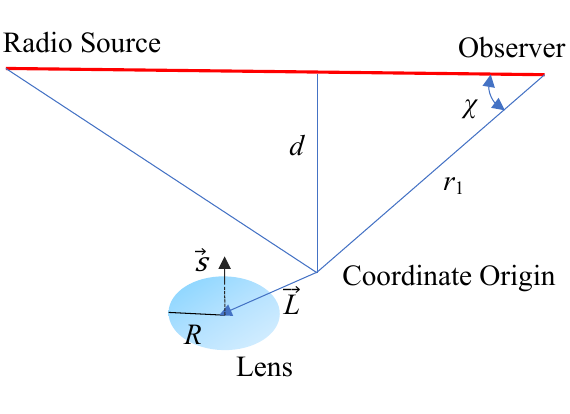}}
    \subfloat[]{\includegraphics[height=0.2\textwidth]{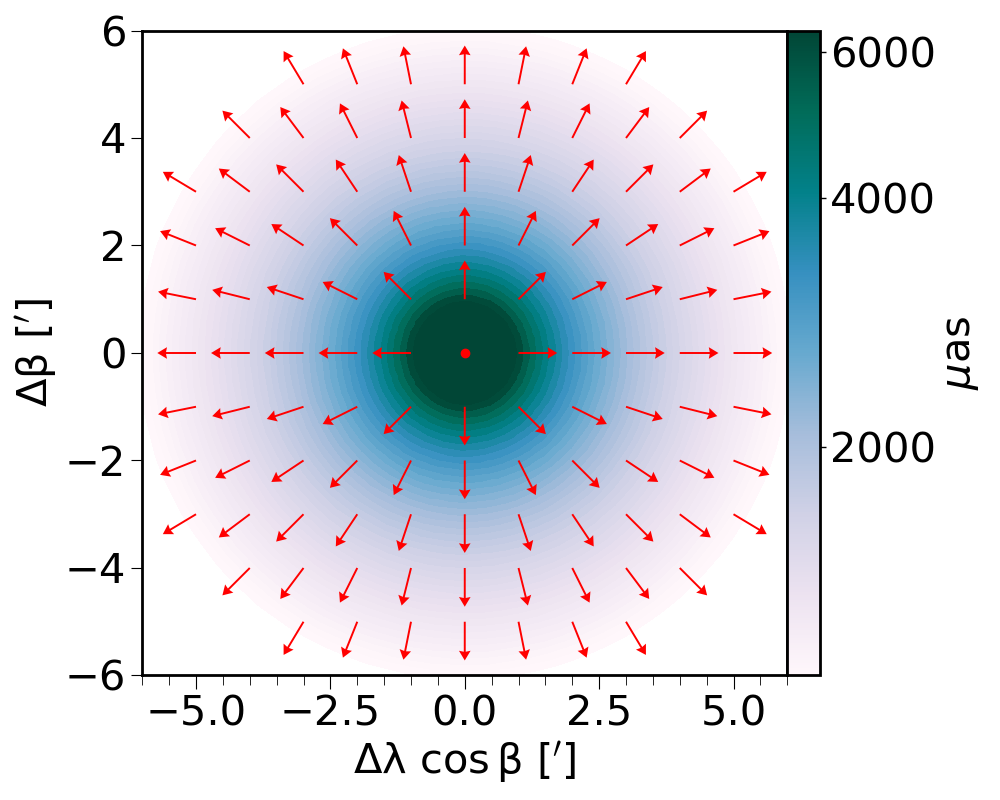}}
    \subfloat[]{\includegraphics[height=0.2\textwidth]{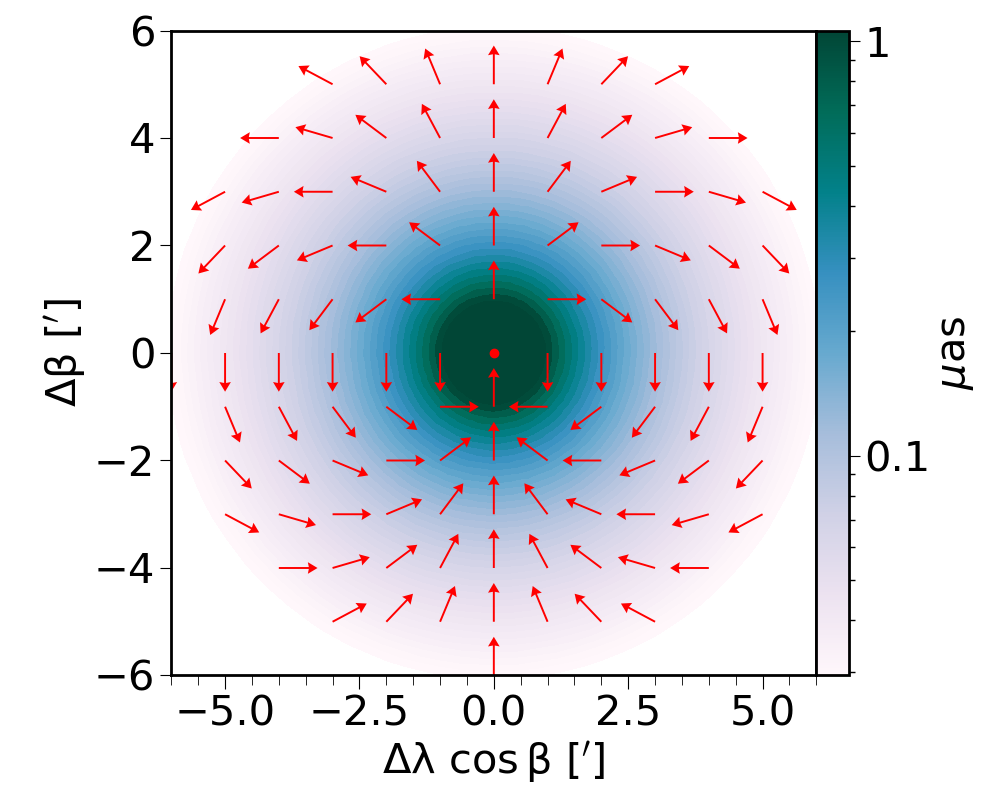}}
    \subfloat[]{\includegraphics[height=0.2\textwidth]{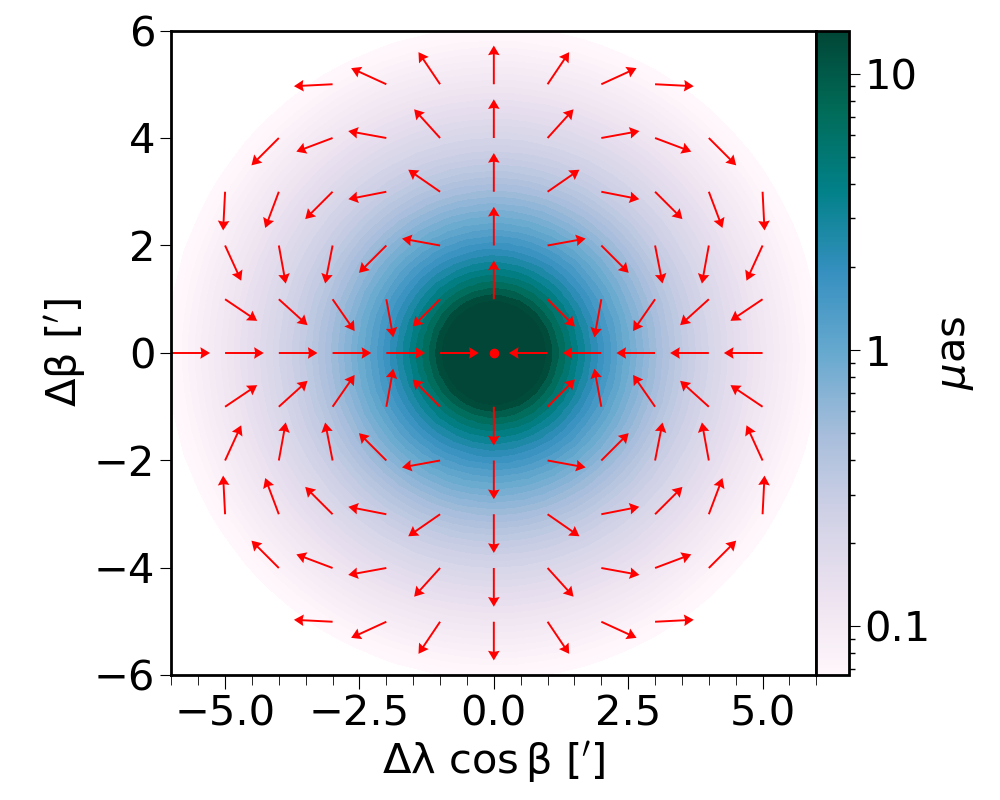}}
    \vspace{-0.2cm}
	\caption{(a) Geometry of light propagation \citep[adapted from][]{Kopeikin+2007}. Light deflection with Jupiter as a lens: (b) monopolar, (c) dipolar from coordinate mis-centering, and (d) quadrupolar from Jupiter's oblateness. }
    \label{fig:geometry-light-deflection}
\end{figure}

The monopolar deflection of light, $\alpha$, can be expressed as \citep[see the geometry of light propagation and light deflection in Figure \ref{fig:geometry-light-deflection};][]{Kopeikin-Makarov2007}
\begin{equation}\label{equ:alpha-mono-light-deflection}
    \alpha = (1 + \gamma)\frac{GM}{c^2d}(1 + \cos \chi)(\vec{e}_{\upbeta} \cos \varphi + \vec{e}_{\uplambda} \sin \varphi),
\end{equation}
where the unit vector $\vec{e}_{\uplambda}$ and $\vec{e}_{\upbeta}$ are mutually orthogonal and directed along the increasing ecliptic longitude, $\uplambda$, and latitude, $\upbeta$, respectively, and $\varphi$ is the position angle of an extragalactic radio source relative to the lens. The ecliptic coordinate system is transformable into the Equatorial or Galactic Coordinate Systems. The impact parameter, $d = r_1 \sin \chi$, represents the minimum approach distance of gravity-unperturbed light to the lens, where $\chi$ denotes the corresponding angular separation and $r_1$ the Earth-lens distance. $M$, $G$, and $c$  represent the lens mass, the gravitational constant, and the speed of light, respectively. 

\begin{figure}[ht]
    \centering
    \subfloat[]{\includegraphics[height=0.37\textwidth]{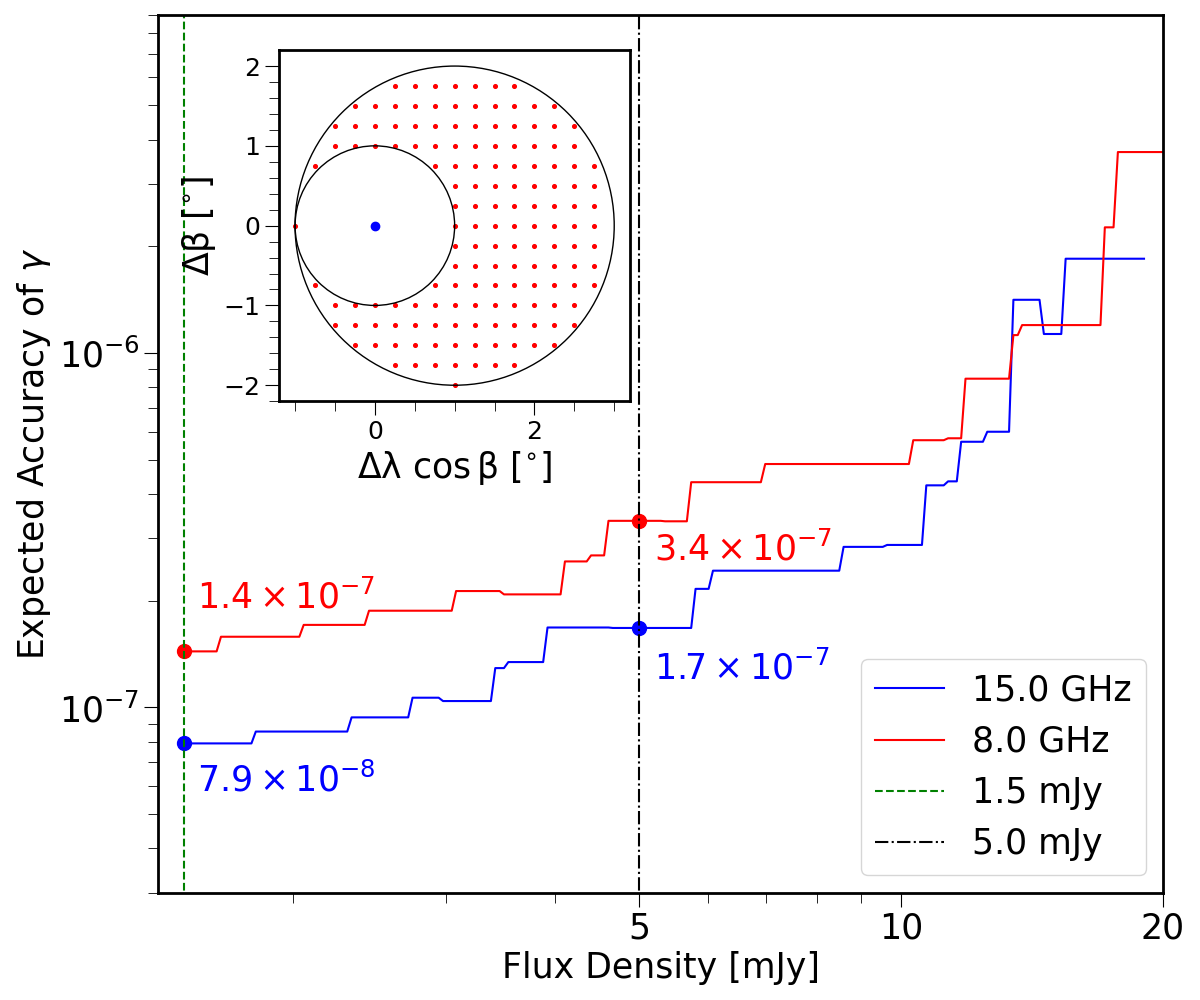}} \hspace{0.5 cm}
    \subfloat[]{\includegraphics[height=0.37\textwidth]{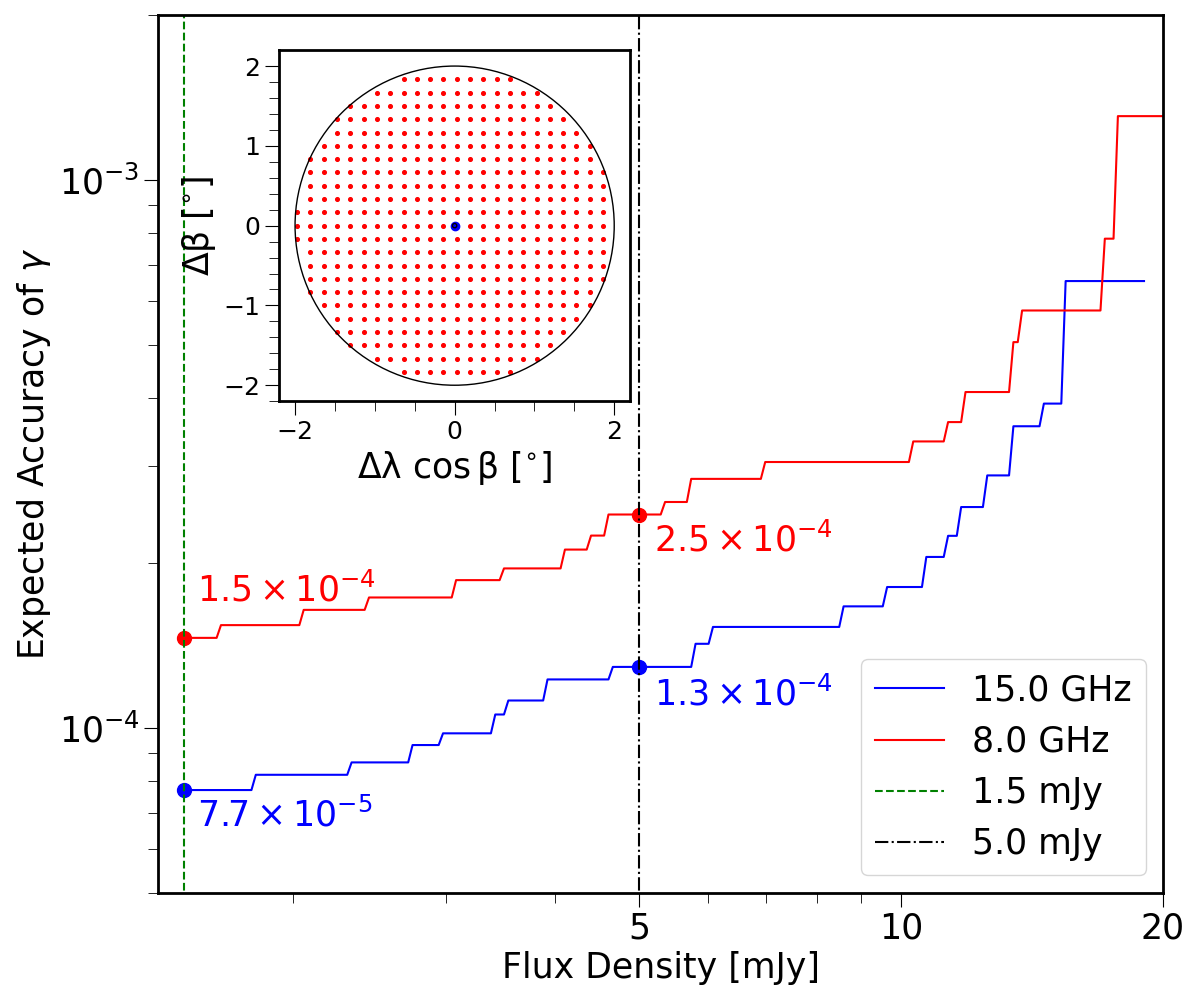}}
    \vspace{-0.2cm}
	\caption{Expected accuracy of $\gamma$ versus flux density for the Sun (a) and Jupiter (b) acting as lenses. Inset in Panel (a): Assumed extragalactic source distribution at 15 GHz and 5.0 mJy for the Sun as a lens. Inset in Panel (b): The same as inset in panel (a) but for Jupiter at 1.5 mJy.}
    \label{fig:gamma_sun_monopolar-light-deflection}
\end{figure}

Following the $\gamma$ measurement methodology of \citet{Li+2022}, observable radio sources were paired, and their relative position variations were simulated over 10 epochs to estimate the per-pair $\gamma$ accuracy, $\sigma_{\gamma_{\mathrm{AB}}}$. This is $\sim$$\Delta \theta/(\alpha_{\mathrm{AB}}\sqrt{n_{\mathrm{epoch}}})$, where $\alpha_{\mathrm{AB}}$ is the $\alpha$ difference within a pair under the assumption of the radio source distribution shown in Figure \ref{fig:gamma_sun_monopolar-light-deflection}, and $n_{\mathrm{epoch}} = 10$. These individual estimates were then combined to determine the overall $\gamma$ accuracy, given by $(\sum \sigma_{\gamma_{\mathrm{AB}}}^{-2})^{-1/2}$.
With a superb sensitivity of $\sim$65 $\mu$Jy (10-sec integration at 15 GHz), SKA-Mid AA$^{\ast}$ can support preliminary light deflection measurements by mapping sources along ephemeris-derived lens trajectories. The experiment will be conducted when the majority of sources are compact extragalactic and in-beam observations are feasible. The $\gamma$ measurement focuses on target sources with flux densities $\geq$1.5 mJy, under which the astrometric precision reaches $\sim$2 and $\sim$4 $\mu$as at 15 and 8 GHz, respectively. With this precision, the accuracy of $\gamma$ under various flux densities is evaluated.

Figure \ref{fig:gamma_sun_monopolar-light-deflection}(a) shows the accuracy estimates of $\gamma$ derived from 10-epoch observations at varying flux densities for the Sun (at its minimum distance of 0.98 au) as a gravitational lens, with 300-sec on-source integration per epoch. A region within $1^{\circ}$ of the Sun is masked to mitigate plasma effects \citep{Titov+2018}. For targets with flux densities $\geq$5 mJy, the integration time can be reduced to $\sim$28 sec to achieve a DR of 100:1. The results demonstrate that the expected accuracy of $\gamma$ reaches $\sim$$10^{-7}$; the estimates for the random and uniform source distributions are consistent within 3$\sigma$. Because the 15 GHz observations offer superior precision, we will focus on them in the subsequent analysis. 

The precise measurement of multipolar deflection of light by Jupiter requires prior determination of monopolar deflection, $\alpha$. Figure \ref{fig:gamma_sun_monopolar-light-deflection}(b) shows $\gamma$ accuracy estimates across flux densities assuming a mean Jovian distance of $\sim$5.2 au, suggesting the achievable accuracy of $\sim$$10^{-4}$.

Note that the expected accuracy of $\gamma$ represents only the order of magnitude. Although the Sun's barycentric motion must be considered for $\gamma$ accuracies at the $10^{-7}$ level \citep[see][]{Kopeikin+2007, Kopeikin2009}, its effect is negligible for the current order-of-magnitude estimate. This is because: (1) $v_{\odot}/c \sim 10^{-4}$, where $v_{\odot}$ is the Sun's barycentric velocity; (2) the relative precision of the magnitude and direction of $v_{\odot}/c$, and that of the Earth-Sun distance, $r_1$, are both $\ll 10^{-3}$; (3) the angular precision of the Sun-Earth and Sun-radio source vectors, each relative to the magnitude of $\chi$, is also $\ll 10^{-3}$. The frequency shift is much smaller than the bandwidth (see Table \ref{tab:SKA-specification-light-deflection}). Lens motion over time increases $\alpha_{\mathrm{AB}}$ for some pairs as the source-lens relative positions change. This does not change the order-of-magnitude estimate. Similarly, the barycentric motion of Jupiter does not affect the order-of-magnitude estimate of the expected $\gamma$ accuracy.

The maximum linearized relativistic time delay, $\Delta \tau$, for a baseline $B=10,000$ km can be expressed as \citep{Fomalont-Kopeikin2003}
\begin{equation}\label{equ:delay-light-deflection}
    \Delta \tau = - \frac{4 G M_{\mathrm{J}}}{c^3 r_{\mathrm{J}}} \left(\frac{B}{\chi} + \frac{3B v_{\mathrm{J}}}{c_{g}\chi^2}\right),
\end{equation}
where Jovian distance to Earth is adopted as $r_{\mathrm{J}}$ = 5.2 au and its velocity as $v_{\mathrm{J}}$ = 17 km s$^{-1}$, and $c_{g}$ and $M_{\mathrm{J}}$ are the retarded deflection and Jovian mass, respectively. With $\chi = 1'$ (defined above), this yields a relative precision of $\sim${}$6 \times 10^{-4}$ for $c_g/c$, 
where the $\Delta \tau$ uncertainty is $\sim$$3 \times 10^{-13}$ s (corresponding to $\Delta \theta \sim 2~\mu$as). Note that there is a controversy over the interpretation of the retarded term \citep[e.g.,][]{Will2003, Fomalont-Kopeikin-2007, Will2014}. SKA-VLBI may provide an opportunity to resolve this controversy.

\subsection{Dipolar and Quadrupolar Light Deflection and Its Application}\label{sec:alpha-high-order-light-deflection}

The multipolar deflections of light caused by the gravitational fields of the Sun and planets, primarily Jupiter and Saturn, are estimated following \citet{Kopeikin-Makarov2007}. We define $\vec{\alpha}_{\mathrm{D}}$ and $\vec{\alpha}_{\mathrm{L}}$ as the dipolar and quadrupolar deflections arising from the displacement of the lens center of mass relative to the coordinate origin, which is set to the ephemeris-derived position. $\vec{\alpha}_{\mathrm{Q}}$ denotes the quadrupolar deflection due to the lens physical oblateness, $J_2$. $\vec{\alpha}_{\mathrm{D}}$, $\vec{\alpha}_{\mathrm{L}}$ and  $\vec{\alpha}_{\mathrm{Q}}$ are given by (see geometry of them in Figure \ref{fig:geometry-light-deflection})
\begin{equation}\label{equ:alpha-D-light-deflection}
    \vec{\alpha}_{\mathrm{D}} = 2(1+\gamma)\frac{GM}{c^2d} \frac{L}{d}[\vec{e}_{\upbeta} \cos (\psi - 2 \varphi) - \vec{e}_{\uplambda} \sin (\psi - 2 \varphi)],
\end{equation}
\begin{equation}\label{equ:alpha-L-light-deflection}
    \vec{\alpha}_{\mathrm{L}} = 2(1+\gamma)\frac{GM}{c^2d} \frac{L^2}{d^2}[\vec{e}_{\upbeta} \cos (2\psi - 3 \varphi) - \vec{e}_{\uplambda} \sin (2\psi - 3\varphi)],
\end{equation}
\begin{equation} \label{equ:alpha-J2-deflection-light-deflection}
    \vec{\alpha}_{\mathrm{Q}} = 2(1+\gamma)\frac{GM}{c^2d} \frac{J_2R^2}{d^2}[\vec{e}_{\upbeta} \cos (2\omega - 3 \varphi) - \vec{e}_{\uplambda} \sin (2\omega - 3\varphi)],
\end{equation}
where $\psi$ denotes the position angle of the vector extending from the coordinate origin to the lens center of mass $\vec{L}$, with magnitude $L$. $\omega$ and $R$ represent the position angle of the lens spin vector $\vec{s}$ and the equatorial radius, respectively. For simplicity, we set $\psi = 0$ and $\omega = 0$.

We model Jupiter as a gravitational lens under two representative scenarios (see Table \ref{tab:alpha_quadrupolar-light-deflection} for Jovian distance $r_1$ and transverse velocity relative to the Earth $v_{\mathrm{p}}$). We assume that Jupiter moves along a straight, ecliptic-parallel path, with a minimum angular separation $\chi_{\mathrm{min}} = 1'$ between Jupiter and the gravity-unperturbed light from the radio source \citep[analogous to those in][]{Kopeikin-Makarov2007}. The positional offset of Jupiter's centre of mass from the coordinate origin is taken as 10,000 $\mu$as, consistent with the ephemerides precision \citep[e.g., JPL DE438 and subsequent ephemerides;][]{Li+2022}.

\begin{table*}[htbp]
	\centering
		\footnotesize
	\setlength\tabcolsep{2.2pt}
	\renewcommand{\arraystretch}{1.1}
	\begin{threeparttable}
		\caption{Workflow of Mesurement of $\vec{\alpha}_{\mathrm{Q}}$ and $J_{\mathrm{2, J}}$\label{tab:alpha_quadrupolar-light-deflection}}
		\begin{tabular}{lcccccccccccccc}
			\hline \hline
			Cases & $r_1$ & $v_p$ & $\chi_{\mathrm{min}}$ & $n_{\mathrm{tot}}$ & $\vec{\alpha}_{\mathrm{D, max}}$ & $\vec{\alpha}_{\mathrm{D, DR}}$ & 
            $\vec{\alpha}_{\mathrm{L, max}}$ & $\vec{\alpha}_{\mathrm{L, DR}}$ &
            $\Delta \theta$ & $\alpha_{\mathrm{cal}}$ &
            $\vec{\alpha}_{\mathrm{Q, max}}$ & $\vec{\alpha}_{\mathrm{Q, DR}}$ & $\sigma_{\vec{\alpha}_{\mathrm{Q}}}/|\vec{\alpha}_{\mathrm{Q}}|$ & $\sigma_{J_{\mathrm{2, J}}}/J_{\mathrm{2, J}}$ \\
            & (au) & (km s$^{-1}$) & (arcmin) & & ($\mu$as) & ($\mu$as) & ($\mu$as) & ($\mu$as) & ($\mu$as) & ($\mu$as) & ($\mu$as) & ($\mu$as) & & \\
			\hline
            case 1 & 4.2 & 17 & 1 & 360 & $\sim$1.1 & $\sim$1.4 & $\ll$0.1 & $\ll$0.1 & $\sim$2 & $\lesssim$0.1 & $\sim$28.7 & $\sim$41.8 & $\sim${}$4\times10^{-3}$ & $\sim${}$4\times10^{-3}$ \\
            case 2 & 6.3 & 43 & 1 & 360 & $\sim$0.7 & $\sim$0.9 & $\ll$0.1 & $\ll$0.1 & $\sim$2 & $\lesssim$0.1 & $\sim$8.5 & $\sim$12.4 & $\sim${}$10^{-2}$ & $\sim${}$10^{-2}$ \\   
			\hline
		\end{tabular}
		\begin{tablenotes}
			\item \parbox{\linewidth}{
                  \setlength{\parindent}{-2em}
                  \setlength{\leftskip}{-1.5em}
                  \setlength{\rightskip}{0pt plus 0pt minus -2em} \noindent\hspace*{-0.4em} \noindent Note. Jupiter and Earth reside on the same side of the Sun for case 1 and on opposite sides for case 2.}
		\end{tablenotes}
	\end{threeparttable}
\end{table*}

\begin{figure}[ht]
    \centering
    \subfloat[]{\includegraphics[height=0.32\textwidth]{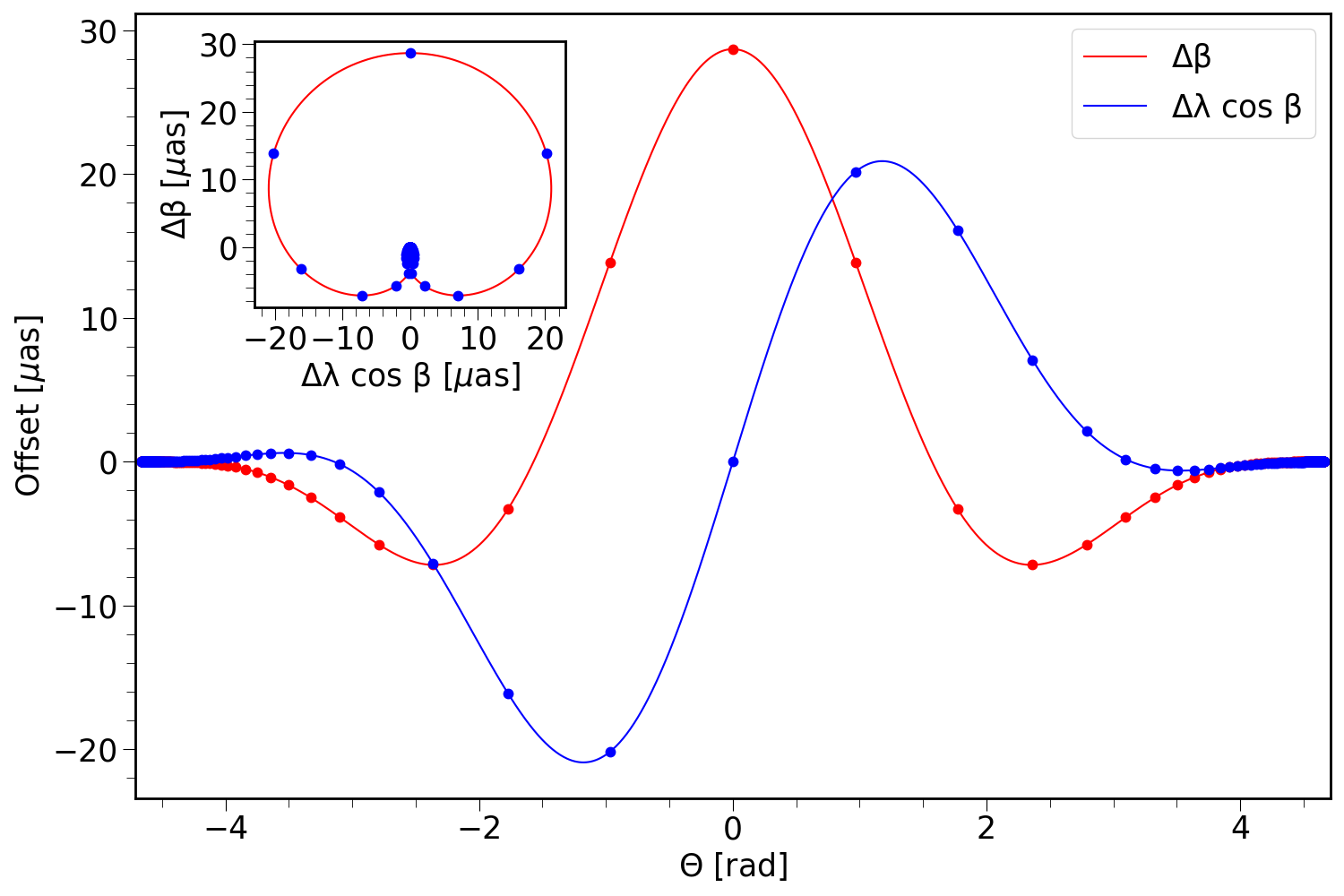}}\hspace{0.2 cm}
    \subfloat[]{\includegraphics[height=0.32\textwidth]{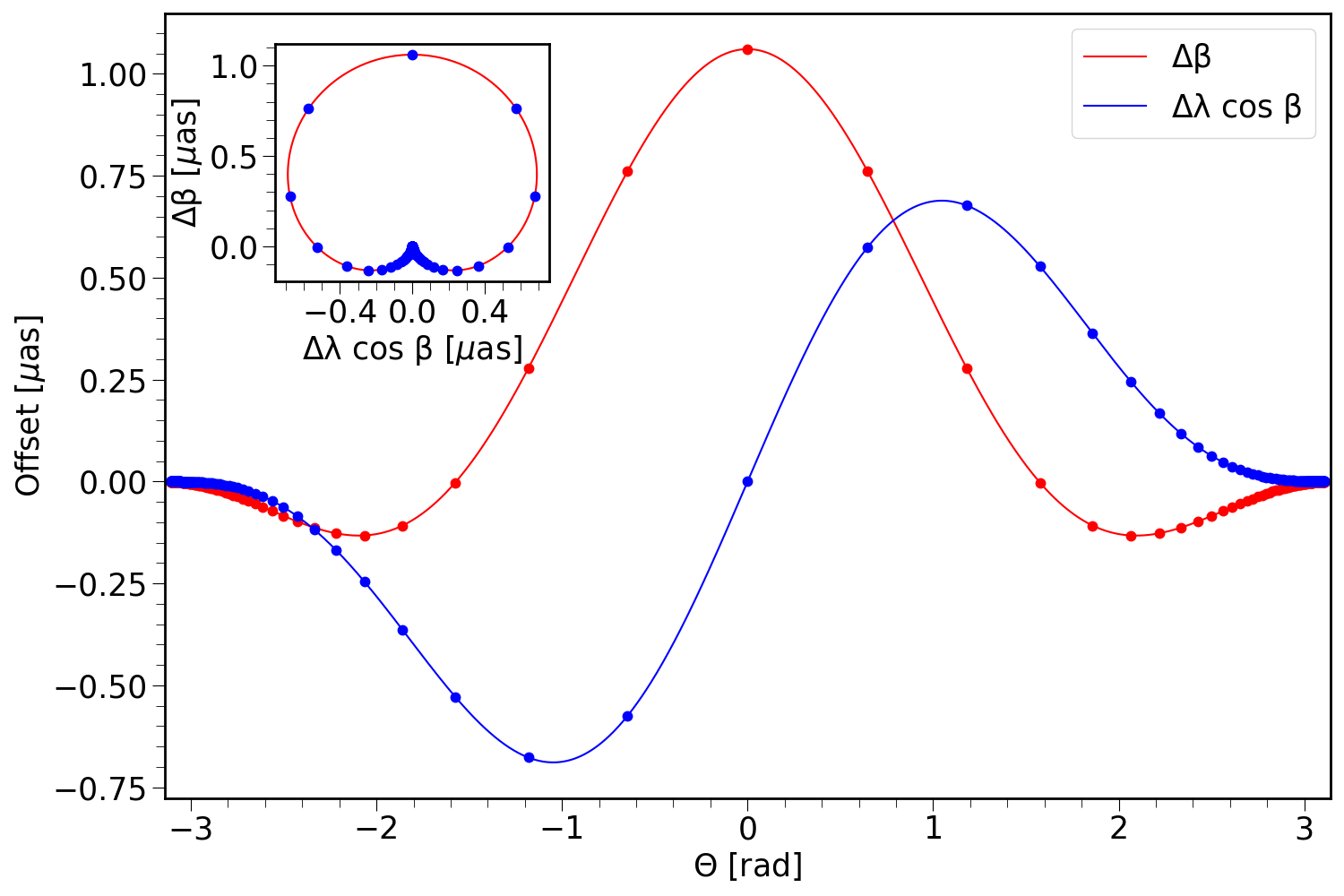}}
    \vspace{-0.2 cm}
	\caption{Variation of $\vec{\alpha}_{\mathrm{Q}}$ (a) and $\vec{\alpha}_{\mathrm{D}}$ (b). Insets display the trajectories of the apparent position of a source deflected by Jupiter. Adjacent points represent 1 hr intervals. $\Theta$ is used for visualisation purposes only.}
    \label{fig:deflection_multipolar-light-deflection}
\end{figure}

The observational sequence initiates with monopolar deflection measurements to ascertain whether any bright extragalactic sources ($\geq$9 mJy at 15 GHz) are within an angular separation of $\chi \leq 1'$ from Jupiter, a threshold determining progression to multipolar deflection measurements. For such sources, the integration time reduces to $\sim$10 sec per sampling point to achieve a DR of 100:1. An assumed campaign comprises 6 epochs (can be selected from the blue 
points in Figure \ref{fig:deflection_multipolar-light-deflection}) with 60 sampling points per epoch, i.e., a total of $n_{\mathrm{tot}} = 360$ samples. Pairing with a secondary calibrator enables monitoring of target-source multipolar variations. For instance, if the calibrator is located 7$'$ away, the total contribution from its dipolar and quadrupolar deflections, $\alpha_{\mathrm{cal}}$, is $\lesssim$0.1 $\mu$as and can thus be ignored.

Figure \ref{fig:deflection_multipolar-light-deflection} shows the variation of $\vec{\alpha}_{\mathrm{D}}$ and $\vec{\alpha}_{\mathrm{Q}}$ during an observational campaign. Table \ref{tab:alpha_quadrupolar-light-deflection} presents the maximum and dynamic range of $\vec{\alpha}_{\mathrm{D}}$, $\vec{\alpha}_{\mathrm{Q}}$, and $\vec{\alpha}_{\mathrm{L}}$, denoted by the subscripts ``max'' and ``DR'', respectively. Treating $\vec{\alpha}_{\mathrm{D, DR}}$ as the uncertainty in measuring $\vec{\alpha}_{\mathrm{Q}}$, and given $\Delta \theta \sim 2$ $\mu$as and $\alpha_{\mathrm{cal}} \lesssim 0.1$ $\mu$as (see Table \ref{tab:alpha_quadrupolar-light-deflection}), $\vec{\alpha}_{\mathrm{Q}}$ becomes measurable for the first time. The relative precision of $\vec{\alpha}_{\mathrm{Q}}$, $\sigma_{\vec{\alpha}_{\mathrm{Q}}}/|\vec{\alpha}_{\mathrm{Q}}|\sim (\Delta\theta + \vec{\alpha}_{\mathrm{D, DR}} + \alpha_{\mathrm{cal}})/(\vec{\alpha}_{\mathrm{Q, DR}} \sqrt{n_{\mathrm{tot}}})$, is $\sim${}$4 \times 10^{-3}$ for case 1 and $\sim${}$10^{-2}$ for case 2. Note that these values are only indicative of the order of magnitude. Reducing $\chi_{\mathrm{min}}$ may further improve the relative precision of $\vec{\alpha}_{\mathrm{Q}}$ because $\vec{\alpha}_{\mathrm{D}} \propto 1/\sin^2 \chi_{\mathrm{min}}$ and $\vec{\alpha}_{\mathrm{Q}} \propto 1/\sin^3 \chi_{\mathrm{min}}$. Measurements of $\vec{\alpha}_{\mathrm{Q}}$ will be conducted at strategically selected epochs during the campaign.

Given that $\gamma$ is determined through measurements of monopolar light deflection, the quadrupolar deflection $\vec{\alpha}_{\mathrm{Q}}$ can be employed to measure Jupiter's physical oblateness, $J_{\mathrm{2, J}}$ (see Equation \eqref{equ:alpha-J2-deflection-light-deflection}). The precision of $J_{\mathrm{2, J}}$ is dominated by that of $\sigma_{\vec{\alpha}_{\mathrm{Q}}}$ because $\gamma$ is measured with higher accuracy. The expected relative precision of $J_{\mathrm{2, J}}$ is $\sigma_{J_{\mathrm{2, J}}}/J_{\mathrm{2, J}} \sim 4 \times 10^{-3}$ for case 1 and reduces to $\sim$0.01 for case 2. $\psi$ or $\omega$ induces rotation of the trajectory in the inset diagram of Figure \ref{fig:deflection_multipolar-light-deflection} while shifting the peak positions in the main figure. The assumption that $\psi = \omega = 0$ does not affect the estimated precision.

Similar calculations were performed for Saturn. Measurement of Saturn's physical oblateness, $J_{\mathrm{2, S}}$, is feasible only when $\chi_{\mathrm{min}} < 30''$ and Saturn lies on the same side of the Sun as Earth (assuming $r_1 = 8.0$ au and $v_{\mathrm{p}} = 20$ km s$^{-1}$). The relative precision of $J_{\mathrm{2, S}}$ will reach $\sim$0.01, corresponding to $\sim$0.01 for $\vec{\alpha}_{\mathrm{Q}}$, with the maximum signal-to-noise ratio limited to $\sim$4 for a single sampling point. For the Sun, the quadrupolar deflection induced by its physical oblateness, $J_{2,\odot}$, remains below 1 $\mu$as, rendering measurement challenges.

In addition, measurements of Mercury's perihelion precession will facilitate the determination of PPN parameter $\beta$ \citep{Will2014, Li+2024-SKA-VLBI}. Mercury's annual precession reads $\Delta \Phi \sim 429.8\times10^3$ $\mu$as$\cdot[(2 + 2\gamma - \beta) + 3 \times 10^{-4} J_{2,\odot}/10^{-7}]$ \citep{Will2014}. Assuming an annual measurement precision of $\Delta \Phi$ at 2 $\mu$as, with $J_{2,\odot} = (2.252 \pm 0.024) \times 10^{-7}$ and $\gamma$ accuracy of $\sim$10$^{-7}$, $\beta$ can be determined to $\sim${}$3 \times 10^{-6}$, which may also provide constraints on other PPN parameters \citep{Will2014}.

Measurements of $\gamma$, or even $\beta$, and the multipolar deflection of light will facilitate rigorous tests and development of GR, the alternative gravitational theories, and high-order PPN formalisms. For instance, achieving an accuracy of $\gamma$ at the $10^{-7}$ level would constrain the Brans-Dicke coupling parameter, $\omega_{\mathrm{BD}}$, in Brans-Dicke theory to $>$10$^{7}$ \citep{Will2014}. In return, applying these advanced gravitational frameworks or high-order PPN formalisms to astrometry, especially in the calibration of geometric delays, provides the theoretical foundation for achieving (sub-)$\mu$as astrometric precision. 

\section{Gravitational versus Plasma Light Deflection: A Black Hole Case Study}\label{sec:others-light-deflection}

The study and application of light deflection, including achromatic gravitational deflection and chromatic deflection, constitutes an interesting astronomical research topic. We employ a black hole paradigm to examine light deflection induced by nearby or massive objects. \citet{Sahu+2025} confirmed the presence of an isolated stellar-mass black hole (the lens) in the microlensing event OGLE-2011-BLG-0462 and yielded the parameters: for the lens itself, a distance of $d_{\mathrm{BH}} \sim 1.52$ kpc and a mass of $M_{\mathrm{BH}} \sim 7.15\; M_{\odot}$; and for the lens-source star pair, an angular separation of $\chi_0 \sim 4\times10^5$ $\mu$as and a relative proper motion of $\mu \sim 7.4\times10^3$ $\mu$as yr$^{-1}$.

The gravitational deflection, $\alpha_{\mathrm{grav}}$, of the source star's light and its 5-year variation, $|\Delta \alpha_{\mathrm{grav}}|$, are
\begin{equation}\label{equ:delta_grav_BH-light-deflection}
    \alpha_{\mathrm{grav}} = \frac{4G M_{\mathrm{BH}}}{c^2 d_{\mathrm{BH}}\chi_0}, \;\;
    |\Delta \alpha_{\mathrm{grav}}| = \frac{4G M_{\mathrm{BH}} \cdot 5\mu}{c^2 d_{\mathrm{BH}}\chi_0(\chi_0 + 5\mu) }.
\end{equation}
The results are $\alpha_{\mathrm{grav}} \sim 95$ $\mu$as and $|\Delta \alpha_{\mathrm{grav}}| 
\sim 8$ $\mu$as. SKA-VLBI's $\sim$2 $\mu$as single-epoch astrometric precision enables detection of gravitational deflection variations caused by the lens, thereby refining determinations of lens mass and other parameters. This demonstrates that SKA-VLBI may add a new dimension --- the direct measurement of light deflection magnitude --- to the investigations of detected microlensing events. Note that $4\times10^5$ $\mu$as $\ll$ 10$'$ (the average separation between extragalactic sources at 1.5 mJy and 15 GHz), making such light deflection events challenging to capture due to their small amplitude, rare occurrence, and resource-intensive nature. However, targeted monitoring of known high-probability objects would be feasible and scientifically valuable.

Chromatic light deflection may coexist with gravitational deflection  \citep[see review by][]{Bisnovatyi-Kogan-Tsupko2017}. For the typical case of inhomogeneous plasmas, the dominant non-gravitational contribution is the refractive correction, $\alpha_{\mathrm{refr}}$. Assuming a spherically symmetric plasma distribution with number density $n(r) = n_0 (r_0/r)^h$ where $r_0 = 1$ au, this correction is given by \citep{Bisnovatyi-Kogan-Tsupko2017}
\begin{equation}\label{equ:delta_refr_BH-light-deflection}
    \alpha_{\mathrm{refr}} = - \frac{4\pi e^2}{m_e(2\pi \nu)^2} n_0 \left(\frac{r_0}{d}\right)^{h} \frac{\sqrt{\pi}\Gamma(h/2+1/2)}{\Gamma(h/2)}, \;\;\; \Gamma(x) = \int_0^{\infty} t^{x-1}e^{-t}dt, \;\;\; d = d_{\mathrm{BH}}\chi_0,
\end{equation}
where $\nu = 15$ GHz is the observation frequency, while $e$ and $m_e$ denote the electron charge and mass, respectively. Analogous to gravitational deflection, we define $\Delta \alpha_{\mathrm{refr}}$ as the variation of $\alpha_{\mathrm{refr}}$ resulting from the increase of $\chi_0$ to $\chi_0 + 5\mu$  after five years.

\citet{Motta+2025} inferred a plasma density of $\sim$500 cm$^{-3}$ in the jet-driven shock front (radius $\sim$0.8 pc) of the black hole X-ray binary GRS 1915+105, with a black hole mass of $\sim$11.4 $M_{\odot}$. Assuming the ambient medium surrounding the black hole candidate in OGLE-2011-BLG-0462 has comparable plasma density, $\Delta \alpha_{\mathrm{refr}}$ reaches $\sim$90 and $\sim$10$^{4}$ $\mu$as for $h$ = 1 and 2, respectively. When a source's initial separation $\chi_0$ is $\geq$1$'$, $\Delta \alpha_{\mathrm{refr}} \lesssim$ 0.1 $\mu$as for $h$ = 1--3 while $\Delta \alpha_{\mathrm{grav}} \ll$ 0.1 $\mu$as, establishing its suitability as a calibrator. Consequently, searches for source pairs to measure light-deflection variations are readily feasible.

The above results constitute a preliminary estimate. Although it is premature to conclude that $|\Delta \alpha_{\mathrm{refr}}| > |\Delta \alpha_{\mathrm{grav}}|$, it suggests that plasma-induced chromatic light deflection could have profound astronomical implications, potentially emerging as an important research focus for SKA-VLBI. Unfortunately, Earth's ionospheric plasma contaminates high-precision astrometric measurements. The standard ionospheric calibration will partially remove chromatic signatures from astrophysical plasmas; therefore, multi-frequency scheduling and joint plasma model fits will likely be necessary. Before pioneering measurements and in-depth studies of plasma-induced light deflection commence, accurate plasma models for Earth, the Sun, Jupiter, and other solar system bodies must be established.

\section{Summary and Conclusion}\label{sec:summary-light-deflection}

This chapter examines the test of gravitational theories by measuring light deflection in the gravitational fields of solar system objects, alongside associated applications and potential breakthroughs. The main conclusions are as follows: 
\begin{itemize}
\item $\gamma$ accuracy can achieve $\sim${}$10^{-7}$ using the Sun as a gravitational lens through monopolar light deflection measurements, and $\sim${}$10^{-4}$ from Jupiter-induced deflection observations.
\item The first direct measurement of Jupiter-induced quadrupolar deflection can constrain Jupiter's physical oblateness, $J_{\mathrm{2, J}}$, to within $\sim${}$10^{-3}$. Saturn's $J_{\mathrm{2, S}}$ may be measurable under optimal conditions, whereas determining the solar $J_{\mathrm{2, \odot}}$ remains a challenge.
\item Plasma-induced light deflection will become a major research focus if and when an accurate model of the local plasma environment is established.
\end{itemize}
SKA-VLBI will advance rigorous testing and development of gravitational theories and high-order PPN formalisms. In return, applying advanced gravitational theories or PPN framework to astrometry can establish the theoretical foundation for achieving (sub)$\mu$as precision.

\vspace{0.8\baselineskip}
\textbf{\large Acknowledgements} 

We thank the anonymous referee for the helpful comments and suggestions. This work is supported by the NSFC Grants Nos. 12203104 and 12403077, the National SKA Program of China (grant No. 2022SKA0120103), and the Key Laboratory for Radio Astronomy. 

\bibliographystyle{abbrvnat-maxbibnames4}
\bibliography{Light_deflection} 

\end{CJK*}
\end{document}